\documentclass{Interspeech2024}

\usepackage{setspace} %
\usepackage{booktabs} %
\usepackage{multirow} %
\usepackage[T1]{fontenc}
\usepackage[labelfont=bf]{caption} %
\usepackage{color}
\usepackage[table]{xcolor}
\usepackage{cite}
\usepackage{enumitem}
\usepackage{adjustbox}

\renewenvironment{thebibliography}[1]{
  \begin{oldthebibliography}{#1}
  \setlength{\itemsep}{-0.1em}
  \setlength{\parskip}{0.em}
}
{
  \end{oldthebibliography}
}

\makeatletter
\def\bstctlcite{\@ifnextchar[{\@bstctlcite}{\@bstctlcite[@auxout]}}
\def\@bstctlcite[#1]#2{\@bsphack
  \@for\@citeb:=#2\do{%
    \edef\@citeb{\expandafter\@firstofone\@citeb}%
    \if@filesw\immediate\write\csname #1\endcsname{\string\citation{\@citeb}}\fi}%
  \@esphack}
\makeatother

\interspeechcameraready

\title{URGENT Challenge: Universality, Robustness, and Generalizability\\For Speech Enhancement}

\name[affiliation={1,2}]{Wangyou}{Zhang}
\name[affiliation={3}]{Robin}{Scheibler}
\name[affiliation={4}]{Kohei}{Saijo}
\name[affiliation={1}]{Samuele}{Cornell}
\name[affiliation={1,2}]{Chenda}{Li}
\name[affiliation={5}]{Zhaoheng}{Ni}
\name[affiliation={5}]{Anurag}{Kumar}
\name[affiliation={6}]{Jan}{Pirklbauer}
\name[affiliation={6}]{Marvin}{Sach}
\name[affiliation={1}]{Shinji}{Watanabe}
\name[affiliation={6}]{Tim}{Fingscheidt}
\name[affiliation={2}]{Yanmin}{Qian}

\address{
  $^1$Carnegie Mellon University, USA
  $^2$Shanghai Jiao Tong University, China
  $^3$LY Corp., Japan
  $^4$Waseda University, Japan
  $^5$Meta, USA
  $^6$Technische Universität Braunschweig, Germany}
\email{wyz-97@sjtu.edu.cn, robin.scheibler@linecorp.com, saijo@pcl.cs.waseda.ac.jp}


\begin{document}
\bstctlcite{IEEEexample:BSTcontrol} %

\maketitle

\begin{abstract}
    The last decade has witnessed significant advancements in deep learning-based speech enhancement (SE). However, most existing SE research has limitations on the coverage of SE sub-tasks, data diversity and amount, and evaluation metrics. To fill this gap and promote research toward universal SE, we establish a new SE challenge, named URGENT, to focus on the universality, robustness, and generalizability of SE.
    We aim to extend the SE definition to cover different sub-tasks to explore the limits of SE models, starting from denoising, dereverberation, bandwidth extension, and declipping.
    A novel framework is proposed to unify all these sub-tasks in a single model, allowing the use of all existing SE approaches. We collected public speech and noise data from different domains to construct diverse evaluation data. Finally, we discuss the insights gained from our preliminary baseline experiments based on both generative and discriminative SE methods with 12 curated metrics.
\end{abstract}

\vspace{-8pt}
\section{Introduction}
\label{sec:intro}
Speech enhancement (SE) is the task of improving a speech signal that has been subject to distortions such as additive noise, acoustic interference, reverberation, or bandwidth limitation.
In recent years, we have witnessed the rapid development of deep learning-based SE techniques, with impressive performance under matched conditions~\cite{TF_GridNet-Wang2023}.
However, most conventional SE approaches focus only on denoising or dereverberation in \emph{a limited range of conditions}, such as single-channel, multi-channel, anechoic, etc.
Usually, they tend to only train and evaluate SE models on one or two common datasets, such as the VoiceBank+DEMAND~\cite{Speech-Valentini-Botinhao2016} and Deep Noise Suppression (DNS) Challenge datasets~\cite{INTERSPEECH2020-Reddy2020,ICASSP2021-Reddy2021,INTERSPEECH2021-Reddy2021,ICASSP2022-Dubey2022,ICASSP-Dubey2023}. The evaluation is often restricted to simulated conditions similar to those of training. This greatly impedes a comprehensive understanding of the generalizability and robustness of SE methods.
In addition, such practice can impact the model design process as it can favor models that are only suitable for limited conditions or have limited capacity to handle more complicated scenarios.

Apart from conventional discriminative methods, generative approaches have also attracted a lot of attention. They are good at handling different distortions with a single model~\cite{Universal-Serra2022,VoiceFixer-Liu2022} and tend to generalize better than discriminative methods~\cite{Conditional-Lu2022}. However, their capability and universality have not yet been fully understood through a comprehensive benchmark.
Meanwhile, recent efforts~\cite{Toward-Zhang2023} have shown the possibility of building a single system to handle various input formats, such as different sampling frequencies and numbers of microphones. However, there \emph{lacks a well-established benchmark} covering a wide range of conditions, and, crucially, \emph{no systematic comparison} has been made yet between state-of-the-art (SOTA) discriminative and generative methods regarding their generalizability.

We believe that the community should focus on this problem urgently. And we propose a new challenge, which is called URGENT, to boost the research on \textbf{U}niversality, \textbf{R}obustness, and \textbf{G}eneralizability for speech \textbf{E}nhanceme\textbf{NT}.
The key contributions and innovations of this challenge are listed below:
\begin{itemize}[wide]
	\item[1)] \emph{Broader definition of the SE task}: In most real-world scenarios, speech is likely to be degraded by several of the distortions mentioned previously, and the recording devices may also vary in the sampling frequency. So, it is important to build a universal SE model that can handle different distortions and input formats.
	Although it is possible to build a separate SE model for each distortion and each input format,
    having a single universal SE model is more efficient and simpler to deploy. Crucially, it can avoid the error propagation that occurs when cascading several specialized models.
	It may also improve the overall performance by sharing knowledge among different sub-tasks, a direction to be explored during this challenge.
    To facilitate this exploration, we further propose a technically novel framework (see Section~\ref{ssec:baseline}) that is general for different SE approaches.
	\item[2)] \emph{Larger scale and more diverse data with training data mandated and limited}: As mentioned above, SE models are often evaluated on fixed, small (e.g., \textasciitilde{}10 h) or medium datasets (e.g., \textasciitilde{}100 h). It is very likely that recent SOTA models heavily overfit these datasets. Furthermore, the test sets associated with these datasets are often matched in terms of speech quality, linguistic content, noise family, and other characteristics. The matter is complicated by the scarcity of truly high-quality, anechoic speech recordings. Large-scale speech datasets are typically recorded with diverse types of equipment under diverse ``sub-optimal'' conditions and are not equalized.
	It is unclear how current SE models can scale with a larger amount of ``sub-optimal'' data and whether they can generalize well to unseen conditions. In this challenge, we aim to explore this aspect with public data. But unlike most earlier challenges, we mandate and limit the (still large amount of) training material, giving us better insights into the actual capabilities of the various network architectures.
	\item[3)] \emph{Extensive evaluation metrics}: Existing challenges often only adopt one or two objective metrics for evaluation, which cannot provide a comprehensive understanding of the SE models.
    For example, some models are trained with a particular evaluation metric (e.g., scale-invariant signal-to-noise ratio~\cite{SDR__Half_baked-LeRoux2019}) discriminatively leading to biased evaluation.
    Beyond a variety of task-dependent intrusive and non-intrusive metrics, as a challenge novelty, we also adopt metrics that are downstream task-independent (e.g., phoneme similarity). This not only fits perfectly to our multi-task challenge, but also promises to allow a better comparison of generative and discriminative SE methods.
\end{itemize}

\begin{figure}[t]
  \centering
  \includegraphics[width=\columnwidth]{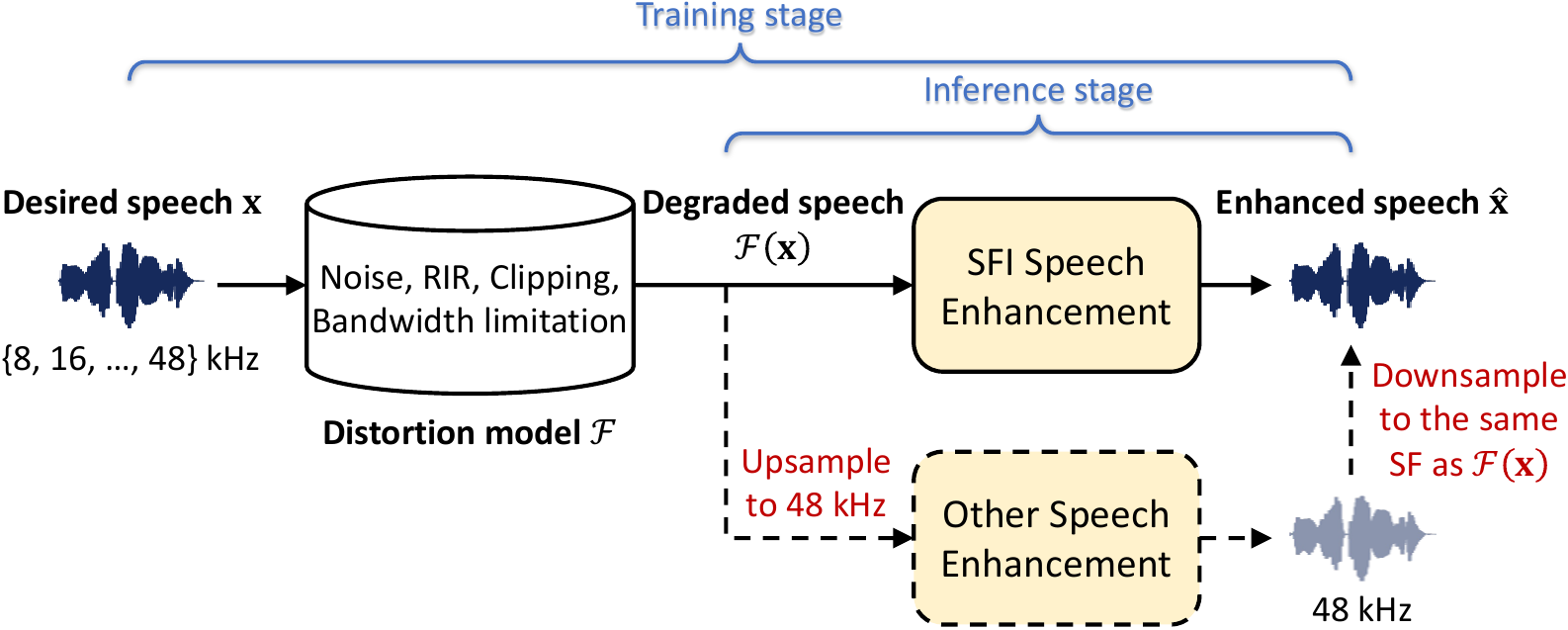}
  \caption{URGENT speech enhancement task definition.}
  \label{fig:overview}
\end{figure}
\begin{figure}[t]
  \centering
  \includegraphics[width=\columnwidth]{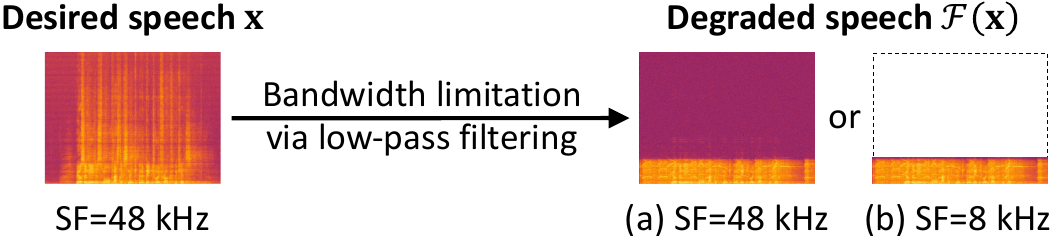}
  \caption{Example of bandwidth limitation (48 kHz $\rightarrow$ 8 kHz).}
  \label{fig:bandwidth}
\end{figure}
\vspace{-5pt}
\section{Related Works}
\label{sec:related}
Existing SE challenges have fostered the development of SE models for specific scenarios, such as denoising and dereverberation~\cite{INTERSPEECH2020-Reddy2020,ICASSP2021-Reddy2021,INTERSPEECH2021-Reddy2021,ICASSP2022-Dubey2022,ICASSP-Dubey2023}, speech restoration~\cite{ICASSP-Cutler2023,ICASSP-Ristea2024}, packet loss concealment~\cite{INTERSPEECH2022-Diener2022}, acoustic echo cancellation~\cite{INTERSPEECH2021-Cutler2021,ICASSP2021-Sridhar2021,ICASSP2022-Cutler2022,ICASSP23AEC-Cutler2023}, hearing aids~\cite{Clarity_2021-Graetzer2021,2nd-Akeroyd2023}, 3D SE~\cite{L3DAS21-Guizzo2021,L3DAS22-Guizzo2022,L3DAS23-Marinoni2024}, far-field multi-channel SE for video conferencing~\cite{ConferencingSpeech-Rao2021}, and unsupervised domain adaptation for denoising~\cite{CHiME_7-Leglaive2023}.
These challenges have greatly advanced SE studies.
The URGENT challenge uniquely focuses on universality, generalizability, and robustness in a wide range of scenarios and evaluation metrics, complementing existing challenges.

\vspace{-8pt}
\section{Challenge Description}
\label{sec:description}

\subsection{Task definition}
\label{ssec:task}
As shown in Figure~\ref{fig:overview}, we define an SE model $\operatorname{SE}(\cdot)$ in the URGENT challenge as the following general form:
{\setlength{\abovedisplayskip}{3pt}
\setlength{\belowdisplayskip}{3pt}
\begin{align}
	\hat{\mathbf{x}} &= \operatorname{SE}(\mathcal{F}(\mathbf{x})) \,, \label{eq:task}
\end{align}
}where $\mathbf{x}$ and $\hat{\mathbf{x}}$ are the desired and enhanced speech signals, respectively. $\mathcal{F}(\cdot)$ is the distortion model that degrades the desired signal.
The resultant degraded speech $\mathcal{F}(\mathbf{x})$ serves as the input to SE models.
Our definition differs from the commonly-adopted definition in the literature in two aspects.
First, the model input $\mathcal{F}(\mathbf{x})$ can have \emph{various sampling frequencies (SF)}, while conventional SE systems often only consider a fixed SF.
Second, the distortion model $\mathcal{F}$ covers \emph{diverse distortions} (i.e., additive noise, reverberation, clipping, and bandwidth limitation), while conventional SE systems mostly apply noise or reverberation.
Note that multichannel signals are not considered in this challenge to simplify the problem.
We leave them for our future challenges.

\vspace{-4pt}
\subsection{Baseline systems unifying various SFs and sub-tasks}
\label{ssec:baseline}
The URGENT challenge prepares several baseline systems, which have been carefully designed to unify different SE sub-tasks in a simple manner.
As shown in Figure~\ref{fig:overview}, the model input $\mathcal{F}(\mathbf{x})$ is simulated by applying different distortions to the original desired speech $\mathbf{x}$.
In this procedure, there are two conditions that involve SF variations and require special treatment.

First, when bandwidth limitation is applied to the desired speech $\mathbf{x}$, corresponding to the bandwidth extension (BWE) sub-task, its high-frequency components are removed via low-pass filtering.
This process typically corresponds to downsampling of the signal, as shown in Figure~\ref{fig:bandwidth} (b).
However, it can also appear in a high SF with its upper frequencies missing due to poor microphone devices, as shown in Figure~\ref{fig:bandwidth} (a).
To unify these two scenarios and to be consistent with other distortions, we always keep the SF unchanged in this procedure as illustrated in Figure~\ref{fig:bandwidth} (a).
We further design the enhanced speech $\hat{\mathbf{x}}$ to have the same SF as input $\mathcal{\mathbf{x}}$ in Figure~\ref{fig:overview} so that we can easily unify the data format in BWE and other SE sub-tasks.

Second, the model input $\mathcal{F}(\mathbf{x})$ can have various SFs as mentioned in Section~\ref{ssec:task}, which cannot be handled by most conventional SE models directly.
One solution is to adopt the so-called sampling-frequency-independent (SFI) SE approaches~\cite{Sampling_frequency_independent-Saito2021,Sampling-Paulus2022,Efficient-Yu2023,Toward-Zhang2023}, as shown in the upper right-hand corner of Figure~\ref{fig:overview}.
The SFI SE models feature a strong generalizability to different SFs, even though only trained on a fixed SF or limited SFs.
Here, we adopt the SFI short-time Fourier transform (STFT) based design due to its zero-shot capability~\cite{Sampling-Paulus2022,Toward-Zhang2023}, which uses \emph{fixed-duration} window and hop sizes in STFT and iSTFT.
Specifically, we apply this design to two recently proposed time-frequency dual-path SE models, e.g., BSRNN\footnote{This differs from BSRNN's original SFI design, where the input signal is always upsampled to 48 kHz. We verified that our design could achieve comparable performance with better generalizability.}~\cite{Efficient-Yu2023} and TF-GridNet~\cite{TF_GridNet-Wang2023} to achieve SFI processing.

On the other hand, we also adopt a simple yet effective solution for most existing SE approaches that only support a single SF~\cite{Perceptually_Motivated-Valin2020}.
As shown in the lower right-hand corner of Figure~\ref{fig:overview}, we always upsample the model input $\mathcal{F}(\mathbf{x})$ to the highest SF (48 kHz) as pre-processing, so the model only takes 48 kHz data as input.
The generated output is also 48 kHz, which will be downsampled to the original input SF for loss calculation as well as for generating the final enhanced speech $\hat{\mathbf{x}}$.
We apply this design to Conv-TasNet~\cite{Conv_TasNet-Luo2019} as an additional baseline.

With this framework (i.e., data and model design), we can easily build an SE system to handle different sub-tasks and SFs.

{
\setlength{\tabcolsep}{3pt}
\begin{table}[t]
    \captionsetup{labelfont=bf}
    \caption{Detailed information of the corpora used in our baseline experiments\protect\footnotemark. $^\dagger$ denotes the data is not used in this paper.}
    \label{tab:corpora}
    \centering
    \resizebox{1.0\columnwidth}{!}{%
        \begin{tabular}{l|ccc} 
        \toprule
        \textbf{Type} & \textbf{Training Set} & \textbf{Validation Set} & \textbf{Non-blind Test Set} \\
        \midrule
        \multirow{5}{*}{Speech} & LibriVox data from DNS5 challenge~\cite{ICASSP-Dubey2023} & \multirow{5}{*}{Same as left} & \multirow{5}{*}{\shortstack[c]{Added one\\unseen corpus}} \\
        & LibriTTS reading speech~\cite{LibriTTS-Zen2019} \\
        & CommonVoice 11.0 English portion~\cite{CommonVoice-Ardila2020}$^\dagger$ \\
        & VCTK reading speech~\cite{VCTK-Veaux2013} \\
        & WSJ reading speech~\cite{WSJ0-LDC1993,WSJ1-Consortium1994}$^\dagger$ \\
        \cellcolor[HTML]{EEEEEE} & \cellcolor[HTML]{EEEEEE}Audioset+FreeSound noise in DNS5 challenge & \cellcolor[HTML]{EEEEEE} & \cellcolor[HTML]{EEEEEE} \\
        \cellcolor[HTML]{EEEEEE}\multirow{-2}{*}{Noise} & \cellcolor[HTML]{EEEEEE}WHAM! noise~\cite{WHAM-Wichern2019} & \cellcolor[HTML]{EEEEEE}\multirow{-2}{*}{Same as left} & \cellcolor[HTML]{EEEEEE}\multirow{-2}{*}{\shortstack[c]{Added two\\unseen corpora}} \\
        \multirow{2}{*}{RIR} & Simulated RIRs from DNS5 challenge & \multirow{2}{*}{Same as left} & \multirow{2}{*}{Real recorded RIRs} \\
        & Other simulated RIRs$^\dagger$ \\
       \bottomrule
       \end{tabular}%
    }
\end{table}
}
\footnotetext{Although both LibriTTS and DNS5 speech data in Table~\ref{tab:corpora} come from LibriVox, they only occupy \textasciitilde{}40\% of the total speech data.}

\vspace{-5pt}
\subsection{Data}
\label{ssec:data}
\vspace{-3pt}
We collect diverse speech, noise, and room impulse response (RIR) samples from public corpora to construct the datasets for this challenge.
As shown in Table~\ref{tab:corpora}, we combine 4 public speech corpora, 2 noise corpora, and 1 RIR corpus for preparing the training and validation sets.
For the non-blind test set, we additionally add 1 unseen speech corpus, 2 unseen noise corpora, and real RIR samples to evaluate the generalizability.
To generate simulation datasets for both training and evaluation, we first preprocess the data as introduced in Section~\ref{sssec:preprocess} and then simulate the data according to Section~\ref{sssec:simulation}.
The data preparation scripts will be made publicly available.

\subsubsection{Preprocessing}
\label{sssec:preprocess}
Since the speech and noise samples are collected from different sources with diverse devices, the effective bandwidth may not be equal to their default SF due to resampling and device discrepancies.
Meanwhile, our baselines in Section~\ref{ssec:baseline} rely on the accurate bandwidth information (ground-truth SF) to perform BWE and other sub-tasks.
And it also allows more accurate metric computation with the actual bandwidth information.
Therefore, it is critical to detect the true bandwidth of each sample and resample it accordingly.
In addition, we observe that speech samples from diverse corpora may be actually non-speech, or can contain noise, or have low quality.
It is thus important to filter out such samples, especially for generative approaches.
To tackle the above issues, we adopt the following procedure as data preprocessing\footnote{The detailed procedure can be found at \url{https://github.com/urgent-challenge/urgent2024_challenge}.}:
\begin{itemize}
	\item[1)] We first follow the algorithm proposed in~\cite{Hi_Fi-Bakhturina2021} to estimate the effective bandwidth of each speech and noise sample, and then resample it to the best matching SF\footnote{The best matching SF is defined as the lowest SF that can fully cover the effective frequency range.}.
	\item[2)] We use a voice activity detection (VAD) algorithm\footnote{\url{https://github.com/wiseman/py-webrtcvad}} to filter out speech samples that are detected to be non-speech or dominated by silence.
	\item[3)] We calculate the DNSMOS scores (OVRL, SIG, BAK)~\cite{DNSMOS-Reddy2022} for each speech sample and set a threshold for each score to filter out noisy and low-quality speech samples.
\end{itemize}
Some interesting observations in this stage are:
\begin{itemize}
	\item While the original LibriVox English data from DNS5 challenge~\cite{ICASSP-Dubey2023} should be in 48 kHz, after the above preprocessing, we found out that about 50\% of them have an SF of 32 kHz, and about 20\% of them have SFs between 8 kHz and 24 kHz. Similar phenomena are also observed in LibriTTS~\cite{LibriTTS-Zen2019} and CommonVoice~\cite{CommonVoice-Ardila2020}.
	\item 19 speech samples in the LibriVox portion of the DNS5 Challenge data are detected to be actually non-speech.
	\item Some speech samples in the LibriVox data from the DNS5 Challenge are found to contain multiple speakers sequentially\footnote{We only detected such samples manually.}. However, we decide not to filter out such samples to allow the SE models to learn to cope with them.
\end{itemize}
Through this process, we finally obtain a curated list of speech (\textasciitilde 1300 hours) and noise (\textasciitilde 250 hours) samples that will be used for data simulation of training and test data in Section~\ref{sssec:simulation}.

\vspace{-6pt}
\subsubsection{Simulation}
\label{sssec:simulation}
We design the data simulation process by considering both speed and reproducibility.
For fixed data simulation, a manifest is firstly generated from the given list of speech, noise, and RIR samples.
It specifies how each sample will be simulated, including the type of distortion to be applied, the speech/noise/RIR sample to be used, the signal-to-noise ratio (SNR), the random seed, and so on.
Then, the simulation can be done in parallel for different samples according to the manifest while ensuring reproducibility.
This procedure can be used to generate training, validation, and non-blind test datasets.
For the training set, we also recommend dynamically generating degraded speech samples during training to increase the data diversity.
It should be noted that only the listed corpora in Table~\ref{tab:corpora} shall be used to generate the training and validation data.
This is to ensure a fair comparison and proper understanding of various SE approaches.
This rule particularly deviates from DNS Challenges which allowed the use of arbitrary training data.

\vspace{-6pt}
\subsection{Evaluation metrics}
\label{ssec:metrics}
To comprehensively evaluate the baseline models, we adopt a wide range of evaluation metrics, including\footnote{The final evaluation metrics in the challenge may differ slightly.}
\begin{itemize}
	\item \emph{intrusive SE metrics}: POLQA~\cite{POLQA-Beerends2013}, PESQ~\cite{PESQ-Rix2001}, extended short-time objective intelligibility (ESTOI)~\cite{ESTOI-Jensen2016}, signal-to-distortion ratio (SDR)~\cite{SDR-Vincent2006}, mel cepstral distortion (MCD)~\cite{MCD-Kubichek1993}, log-spectral distance (LSD)~\cite{LSD-Gray1976};
	\item \emph{non-intrusive SE metrics}: DNSMOS~\cite{DNSMOS-Reddy2022}, NISQA~\cite{NISQA-Mittag2021};
	\item \emph{downstream-task-independent metrics}: phoneme similarity (PhnSim, equal to ``1-LPD'' in~\cite{Evaluation-Pirklbauer2023}), SpeechBERTScore~\cite{SpeechBERTScore-Saeki2024};
	\item \emph{downstream-task-dependent metrics}: speaker similarity (SpkSim), word accuracy (WAcc)\footnote{WAcc is equal to $1-$ word error rate (WER).}.
\end{itemize}
Among them, a lower value in MCD and LSDindicates better performance, while in all other metrics a higher value corresponds to better performance.
The intrusive SE metrics require well-aligned reference speech for calculation, and can reflect the objective quality of enhanced speech.
The non-intrusive SE metrics are calculated by pre-trained neural networks which do not require reference speech. They are useful to evaluate the generative approaches or when no aligned reference speech is available.
The downstream task independent metrics compare the enhanced speech and reference speech based on some task-agnostic representation (e.g., phoneme prediction and discrete tokens).
Note that although they require reference speech as an additional input, no strict alignment is needed.
The PhnSim metric captures frame-wise phone information in the enhanced speech, which is useful for comparing generative and discriminative approaches in the correctness of their generated contents.
The SpeechBERTScore metric measures the similarity between semantic embeddings of reference and enhanced speech.
The downstream task related metrics use a pre-trained model to evaluate the downstream task performance such as speaker similarity and WAcc.
These allow us to further exploit real-recorded data for evaluation.
We use the RawNet3~\cite{Pushing-Jung2022} model pre-trained on VoxCeleb datasets for cosine-based speaker similarity calculation and the OWSM v3.1~\cite{OWSMv3.1-Peng2024} model for WAcc calculation.

The rank of different SE systems will be obtained by considering all these metrics. More specific ranking rules will be updated on the challenge website \url{https://urgent-challenge.github.io/urgent2024/}, which are facilitated by our investigation in this paper.

\begin{table*}
    \caption{Evaluation on non-blind test data.   Results with $^*$ are not fully comparable due to different data and training setups.}
    \label{tab:exp}
    \centering
    \setlength{\tabcolsep}{3pt}
    \resizebox{\textwidth}{!}{%
    \begin{tabular}{lcc|cc|cccccc|cc|cc}
        \toprule
        \multirow{2}{*}{\textbf{Model}} & \multirow{2}{*}{\textbf{\#Param}} & \multirow{2}{*}{\shortstack[c]{\textbf{\#MACs}\\(48 kHz)}} & \multicolumn{2}{c|}{\emph{Non-intrusive SE metrics}} & \multicolumn{6}{c|}{\emph{Intrusive SE metrics}} & \multicolumn{2}{c|}{\emph{Downstream-task-independent}} & \multicolumn{2}{c}{\emph{Downstream-task-dependent}} \\
        & & & \textbf{DNSMOS} $\uparrow$ & \textbf{NISQA} $\uparrow$ & \textbf{POLQA} $\uparrow$ & \textbf{PESQ} $\uparrow$ & \textbf{ESTOI} ($\times 100$) $\uparrow$ & \textbf{SDR (dB)} $\uparrow$ & \textbf{MCD} $\downarrow$ & \textbf{LSD} $\downarrow$ & \textbf{SpeechBERTScore} $\uparrow$ & \textbf{PhnSim} $\uparrow$ & \textbf{SpkSim} $\uparrow$ & \textbf{WAcc} (\%) $\uparrow$ \\
        \midrule
        Noisy input & - & - & 1.64 & 1.76 & 2.50 & 1.63 & 70.40 & 6.11 & 6.76 & 3.99 & \textbf{0.87} & 0.68 & 0.72 & 82.18 \\
        OM-LSA~\cite{Speech-Cohen2001} & - & - & 2.19 & 2.09 & 2.37 & 1.81 & 70.24 & 10.88 & 5.26 & 3.64 & 0.85 & 0.71 & 0.65 & 78.61 \\
        \hline
        VoiceFixer~\cite{VoiceFixer-Liu2022}$^*$ & 116.8 M & - & \textbf{2.93} & \textbf{3.65} & 1.97 & 1.50 & 52.71 & -9.59 & 9.16 & 7.54 & 0.81 & 0.59 & 0.54 & 66.19 \\
        \hline
        Conv-TasNet & 40.0 M & 38 G/s & 2.31 & 2.71 & 3.12 & 2.42 & 79.91 & 14.42 & 3.23 & 2.73 & 0.85 & 0.73 & 0.70 & 76.82 \\
        BSRNN & 37.8 M & 78 G/s & 2.41 & 3.05 & 3.49 & 2.66 & 83.29 & 14.89 & 2.75 & 2.66 & \textbf{0.87} & 0.80 & 0.77 & 82.53 \\
        TF-GridNet & 8.5 M & 401 G/s & 2.43 & 3.06 & \textbf{3.54} & \textbf{2.76} & \textbf{84.05} & \textbf{15.42} & \textbf{2.70} & \textbf{2.39} & \textbf{0.87} & \textbf{0.81} & \textbf{0.78} & \textbf{82.87} \\
        \bottomrule
    \end{tabular}%
    }
    \vspace{-12pt}
\end{table*}

\vspace{-6pt}
\section{Experiments}
\label{sec:exp}
We report hereafter a preliminary investigation on different baselines as mentioned in Section~\ref{ssec:baseline} for the challenge.
This includes comparing generative and discriminative methods using a wide range of metrics introduced in Section~\ref{ssec:metrics}.

\vspace{-6pt}
\subsection{Data configuration}
\label{ssec:exp_data}

As a preliminary investigation before the challenge, we primarily conducted experiments on a fixed simulation dataset to compare different baseline approaches.
The fixed simulation dataset is generated following the procedure described in Section~\ref{sssec:simulation}, resulting in \textasciitilde{}400 hours of training samples, \textasciitilde{}30 hours of validation samples, and \textasciitilde{}15 hours of test samples.
Note that this data is only used for our preliminary exploration, while the final challenge data may be updated according to our findings.
The SNR ranges from -5 dB to 20 dB, and reverberation is added to each sample with a probability of 0.5.
Note that the generated dataset covers a wide range of sampling frequencies, i.e., $\{8, 16, 22.05, 24, 32, 44.1, 48\}$ kHz.
During training, we always segment each sample into 4\,s chunks for better efficiency.
All models are trained using the L1-based time-domain plus frequency-domain multi-resolution loss~\cite{Towards-Lu2022}, where we adopt four STFT window sizes $\{256, 512, 768, 1024\}$ to obtain different time-frequency resolutions.
All baseline experiments have been done using the ESPnet~\cite{ESPnet_SE-Li2021} toolkit.

\subsection{Model configuration}
\label{ssec:exp_model}
As mentioned in Section~\ref{ssec:baseline}, we evaluate the performance of four different baseline models, including BSRNN, TF-GridNet, and Conv-TasNet.%
We follow the best model configuration in the original papers for TF-GridNet and Conv-TasNet, except that the encoder/decoder kernel sizes are scaled to match a 48 kHz input.
In BSRNN, the STFT window and hop sizes are set to 20 ms and 10 ms, respectively.
We stack 6 BSRNN blocks with a relatively large embedding dimension (196) to enhance the model capacity.
Due to the space limitation, we omit the details for the model configuration, which can be found in the official repository\footnote{\url{https://github.com/urgent-challenge/urgent2024_challenge}} for reproducibility.

\begin{figure}[t]
  \centering
  \includegraphics[width=0.9\columnwidth]{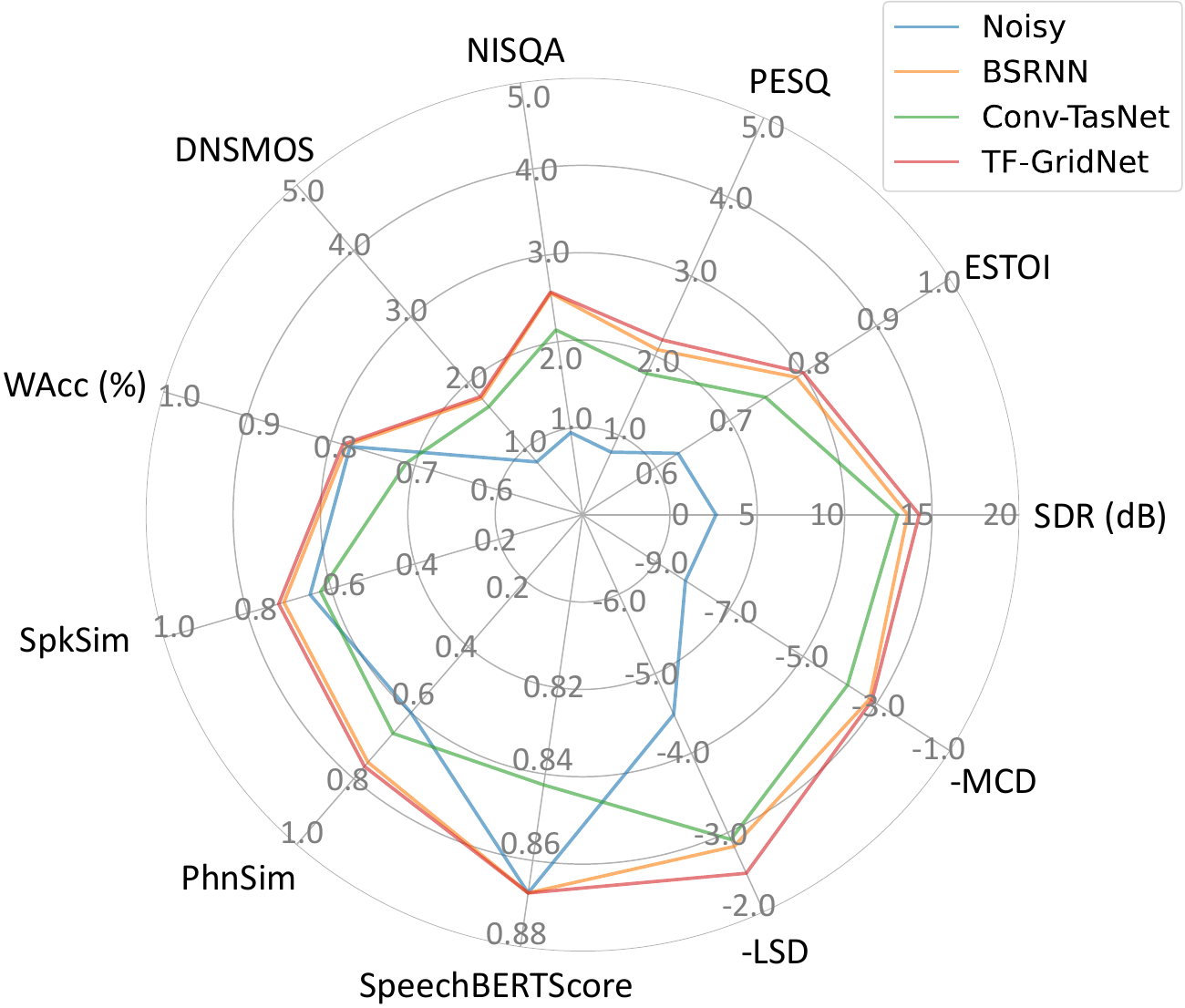}
  \caption{Radar plot of different baseline models.}
  \label{fig:radar}
\end{figure}
\vspace{-4pt}
\subsection{Experimental results and discussion}
\label{ssec:exp}
As shown in Table~\ref{tab:exp}, we compare the performance of different SE models with a wide range of evaluation metrics on the simulated non-blind test set.
Our baselines trained on the challenge data show consistent improvement in most metrics.
Among them, the masking-based SE approach (Conv-TasNet) has the worst performance, which is clearly illustrated in Figure~\ref{fig:radar}.
The breakdown results imply that it cannot work well on band-limited samples, which is attributed to the inherent limitation of masking-based approaches\footnote{We provide the breakdown results in Table~\ref{tab:breakdown} in the Appendix.}.
In contrast, mapping-based methods (i.e., BSRNN and TF-GridNet) show better performance in all SE sub-tasks, demonstrating their potential for unifying multiple SE sub-tasks.
It is also interesting that all metrics share a similar tendency among the discriminative approaches.
This verifies the feasibility of building a universal SE system with a high-capacity SE model.%

In addition to the baseline models described in Section~\ref{ssec:baseline}, we also present the evaluation results of OM-LSA~\cite{Speech-Cohen2001} and VoiceFixer~\cite{VoiceFixer-Liu2022}.
The former is a representative denoising method based on signal processing, which serves as a weak baseline since it can only handle the denoising sub-task.
VoiceFixer is a vocoder-based generative SE approach\footnote{Available at \url{https://github.com/haoheliu/voicefixer}. We adopted ``mode 0'' as it performs best.}, which is trained to process the same set of distortions as mentioned in Section~\ref{ssec:task}.
Note that VoiceFixer is trained to only process data in 44.1 kHz, so we always resample the input to 44.1 kHz and the output back to the original SF for evaluation.
Since VoiceFixer is trained on a different dataset, no fair comparison can be made.
Thus, this model only serves as a reference to check the effectiveness of the other baselines.
We leave a comparable generative baseline for future work.
As expected, it achieves unmatched DNSMOS and NISQA scores, confirming the strength of generative SE approaches to generate natural speech.
Meanwhile, our diverse metrics also demonstrate their capability of detecting the ``hallucination'' of generative SE approaches.
For example, the low PhnSim and SpkSim scores can indicate inconsistent contents and speaker traits in the generated speech.
This also verifies the necessity of using a wide range of evaluation metrics to capture various properties of discriminative and generative SE methods.

\vspace{-6pt}
\section{Conclusion}
\label{sec:conclusion}
\vspace{-3pt}
In this paper, we have introduced a new SE challenge, URGENT, which aims to promote research towards universal SE with strong generalizability and robustness.
This new challenge features a broad definition of the SE task, large scale and diverse data based on public corpora, and extensive evaluation metrics.
A novel framework has been proposed to facilitate this exploration, allowing easy extension of existing SE models to handle multiple SE sub-tasks and different sampling frequencies.
We open source all scripts for data preparation, baseline training, and extensive evaluation.
As a preliminary investigation, we conducted experiments with several baselines on the simulated data.
The results verified the potential of both generative and discriminative SE approaches, each dominating a different set of metrics.
Our goal is to attract more research towards building universal SE models with strong robustness and good generalizability.
In future work, we aim to extend the challenge to cover more scenarios, such as additional distortions, more microphone channels, multiple speakers, etc.

\ifinterspeechfinal
\section{Acknowledgment}
\vspace{-4pt}
The experiments were done using the PSC Bridges2 system via ACCESS allocation CIS210014, supported by National Science Foundation grants \#2138259, \#2138286, \#2138307, \#2137603, and \#2138296.
S. Cornell was supported by IC Postdoctoral Research Fellowship Program
at Carnegie Mellon University via ORISE. 
\fi
\vspace{-6pt}
\bibliographystyle{IEEEtran}
\bibliography{mybib}

\begin{thebibliography}{10}
\providecommand{\url}[1]{#1}
\csname url@samestyle\endcsname
\providecommand{\newblock}{\relax}
\providecommand{\bibinfo}[2]{#2}
\providecommand{\BIBentrySTDinterwordspacing}{\spaceskip=0pt\relax}
\providecommand{\BIBentryALTinterwordstretchfactor}{4}
\providecommand{\BIBentryALTinterwordspacing}{\spaceskip=\fontdimen2\font plus
\BIBentryALTinterwordstretchfactor\fontdimen3\font minus
  \fontdimen4\font\relax}
\providecommand{\BIBforeignlanguage}[2]{{%
\expandafter\ifx\csname l@#1\endcsname\relax
\typeout{** WARNING: IEEEtran.bst: No hyphenation pattern has been}%
\typeout{** loaded for the language `#1'. Using the pattern for}%
\typeout{** the default language instead.}%
\else
\language=\csname l@#1\endcsname
\fi
#2}}
\providecommand{\BIBdecl}{\relax}
\BIBdecl

\bibitem{TF_GridNet-Wang2023}
Z.-Q. Wang \emph{et~al.}, ``{TF}-{G}rid{N}et: Integrating full-and sub-band
  modeling for speech separation,'' \emph{IEEE/ACM Trans. ASLP.}, vol.~31, pp.
  3221--3236, 2023.

\bibitem{Speech-Valentini-Botinhao2016}
C.~Valentini-Botinhao \emph{et~al.}, ``Speech enhancement for a noise-robust
  text-to-speech synthesis system using deep recurrent neural networks,'' in
  \emph{Proc. Interspeech}, 2016, pp. 352--356.

\bibitem{INTERSPEECH2020-Reddy2020}
C.~K. Reddy \emph{et~al.}, ``The {INTERSPEECH} 2020 deep noise suppression
  challenge: Datasets, subjective testing framework, and challenge results,''
  in \emph{Proc. Interspeech}, 2020, pp. 2492--2496.

\bibitem{ICASSP2021-Reddy2021}
C.~K. Reddy \emph{et~al.}, ``{ICASSP} 2021 deep noise suppression challenge,''
  in \emph{Proc. IEEE ICASSP}, 2021, pp. 6623--6627.

\bibitem{INTERSPEECH2021-Reddy2021}
C.~K. Reddy \emph{et~al.}, ``{INTERSPEECH} 2021 deep noise suppression
  challenge,'' in \emph{Proc. Interspeech}, 2021, pp. 2796--2800.

\bibitem{ICASSP2022-Dubey2022}
H.~Dubey \emph{et~al.}, ``{ICASSP} 2022 deep noise suppression challenge,'' in
  \emph{Proc. IEEE ICASSP}, 2022, pp. 9271--9275.

\bibitem{ICASSP-Dubey2023}
H.~Dubey \emph{et~al.}, ``{ICASSP} 2023 deep noise suppression challenge,''
  \emph{IEEE Open Journal of Signal Processing}, pp. 1--13, 2024.

\bibitem{Universal-Serra2022}
J.~Serr{\`a} \emph{et~al.}, ``Universal speech enhancement with score-based
  diffusion,'' \emph{arXiv preprint arXiv:2206.03065}, 2022.

\bibitem{VoiceFixer-Liu2022}
H.~Liu \emph{et~al.}, ``{VoiceFixer}: A unified framework for high-fidelity
  speech restoration,'' in \emph{Proc. Interspeech}, 2022, pp. 4232--4236.

\bibitem{Conditional-Lu2022}
Y.-J. Lu \emph{et~al.}, ``Conditional diffusion probabilistic model for speech
  enhancement,'' in \emph{Proc. IEEE ICASSP}, 2022, pp. 7402--7406.

\bibitem{Toward-Zhang2023}
W.~Zhang \emph{et~al.}, ``Toward universal speech enhancement for diverse input
  conditions,'' in \emph{Proc. IEEE ASRU}, 2023.

\bibitem{SDR__Half_baked-LeRoux2019}
J.~Le~Roux \emph{et~al.}, ``{SDR}--half-baked or well done?'' in \emph{Proc.
  IEEE ICASSP}, 2019, pp. 626--630.

\bibitem{ICASSP-Cutler2023}
R.~Cutler \emph{et~al.}, ``{ICASSP} 2023 speech signal improvement challenge,''
  \emph{IEEE Open Journal of Signal Processing}, pp. 1--12, 2024.

\bibitem{ICASSP-Ristea2024}
N.~C. Ristea \emph{et~al.}, ``{ICASSP} 2024 speech signal improvement
  challenge,'' \emph{arXiv preprint arXiv:2401.14444}, 2024.

\bibitem{INTERSPEECH2022-Diener2022}
L.~Diener \emph{et~al.}, ``{INTERSPEECH} 2022 audio deep packet loss
  concealment challenge,'' in \emph{Proc. Interspeech}, 2022, pp. 580--584.

\bibitem{INTERSPEECH2021-Cutler2021}
R.~Cutler \emph{et~al.}, ``{INTERSPEECH} 2021 acoustic echo cancellation
  challenge,'' in \emph{Proc. Interspeech}, 2021, pp. 4748--4752.

\bibitem{ICASSP2021-Sridhar2021}
K.~Sridhar \emph{et~al.}, ``{ICASSP} 2021 acoustic echo cancellation challenge:
  Datasets, testing framework, and results,'' in \emph{Proc. IEEE ICASSP},
  2021, pp. 151--155.

\bibitem{ICASSP2022-Cutler2022}
R.~Cutler \emph{et~al.}, ``{ICASSP} 2022 acoustic echo cancellation
  challenge,'' in \emph{Proc. IEEE ICASSP}, 2022, pp. 9107--9111.

\bibitem{ICASSP23AEC-Cutler2023}
R.~Cutler \emph{et~al.}, ``{ICASSP} 2023 acoustic echo cancellation
  challenge,'' \emph{IEEE Open Journal of Signal Processing}, 2024.

\bibitem{Clarity_2021-Graetzer2021}
S.~Graetzer \emph{et~al.}, ``Clarity-2021 challenges: Machine learning
  challenges for advancing hearing aid processing,'' in \emph{Proc.
  Interspeech}, 2021, pp. 686--690.

\bibitem{2nd-Akeroyd2023}
M.~A. Akeroyd \emph{et~al.}, ``The 2nd clarity enhancement challenge for
  hearing aid speech intelligibility enhancement: Overview and outcomes,'' in
  \emph{Proc. IEEE ICASSP}, 2023.

\bibitem{L3DAS21-Guizzo2021}
E.~Guizzo \emph{et~al.}, ``{L3DAS21} challenge: Machine learning for {3D} audio
  signal processing,'' in \emph{Proc. MLSP}.\hskip 1em plus 0.5em minus
  0.4em\relax IEEE, 2021, pp. 1--6.

\bibitem{L3DAS22-Guizzo2022}
E.~Guizzo \emph{et~al.}, ``{L3DAS22} challenge: Learning {3D} audio sources in
  a real office environment,'' in \emph{Proc. IEEE ICASSP}, 2022, pp.
  9186--9190.

\bibitem{L3DAS23-Marinoni2024}
C.~Marinoni \emph{et~al.}, ``Overview of the {L3DAS23} challenge on
  audio-visual extended reality,'' in \emph{Proc. IEEE ICASSP}, 2023, pp. 1--2.

\bibitem{ConferencingSpeech-Rao2021}
W.~Rao \emph{et~al.}, ``{ConferencingSpeech} challenge: Towards far-field
  multi-channel speech enhancement for video conferencing,'' in \emph{Proc.
  IEEE ASRU}, 2021, pp. 679--686.

\bibitem{CHiME_7-Leglaive2023}
S.~Leglaive \emph{et~al.}, ``The {CHiME}-7 {UDASE} task: Unsupervised domain
  adaptation for conversational speech enhancement,'' in \emph{Proc. CHiME},
  2023.

\bibitem{Sampling_frequency_independent-Saito2021}
K.~Saito \emph{et~al.}, ``Sampling-frequency-independent audio source
  separation using convolution layer based on impulse invariant method,'' in
  \emph{Proc. EUSIPCO}, 2021, pp. 321--325.

\bibitem{Sampling-Paulus2022}
J.~Paulus and M.~Torcoli, ``Sampling frequency independent dialogue
  separation,'' in \emph{Proc. EUSIPCO}, 2022, pp. 160--164.

\bibitem{Efficient-Yu2023}
J.~Yu and Y.~Luo, ``Efficient monaural speech enhancement with universal sample
  rate band-split {RNN},'' in \emph{Proc. IEEE ICASSP}, 2023.

\bibitem{Perceptually_Motivated-Valin2020}
J.-M. Valin \emph{et~al.}, ``A perceptually-motivated approach for
  low-complexity, real-time enhancement of fullband speech,'' in \emph{Proc.
  Interspeech}, 2020, pp. 2482--2486.

\bibitem{Conv_TasNet-Luo2019}
Y.~Luo and N.~Mesgarani, ``{Conv-TasNet}: Surpassing ideal time--frequency
  magnitude masking for speech separation,'' \emph{IEEE/ACM Trans. ASLP.},
  vol.~27, no.~8, pp. 1256--1266, 2019.

\bibitem{LibriTTS-Zen2019}
H.~Zen \emph{et~al.}, ``{LibriTTS}: A corpus derived from {LibriSpeech} for
  text-to-speech,'' in \emph{Proc. Interspeech}, 2019, pp. 1526--1530.

\bibitem{CommonVoice-Ardila2020}
R.~Ardila \emph{et~al.}, ``Common voice: A massively-multilingual speech
  corpus,'' in \emph{Proceedings of the 12th Language Resources and Evaluation
  Conference}, 2020, pp. 4218--4222.

\bibitem{VCTK-Veaux2013}
C.~Veaux, J.~Yamagishi, and S.~King, ``The {Voice} {Bank} corpus: Design,
  collection and data analysis of a large regional accent speech database,'' in
  \emph{Proc. O-COCOSDA/CASLRE}, 2013, pp. 1--4.

\bibitem{WSJ0-LDC1993}
LDC, \emph{{LDC} Catalog: CSR-I (WSJ0) Complete}, University of Pennsylvania,
  1993.

\bibitem{WSJ1-Consortium1994}
{Philadelphia: Linguistic Data Consortium}, \emph{{LDC} Catalog: CSR-II (WSJ1)
  Complete LDC94S13A}, 1994.

\bibitem{WHAM-Wichern2019}
G.~Wichern \emph{et~al.}, ``{WHAM!}: Extending speech separation to noisy
  environments,'' in \emph{Proc. Interspeech}, 2019, pp. 1368--1372.

\bibitem{Hi_Fi-Bakhturina2021}
E.~Bakhturina \emph{et~al.}, ``{Hi-Fi} multi-speaker {English} {TTS} dataset,''
  in \emph{Proc. Interspeech}, 2021, pp. 2776--2780.

\bibitem{DNSMOS-Reddy2022}
C.~K. Reddy, V.~Gopal, and R.~Cutler, ``{DNSMOS} {P}.835: A non-intrusive
  perceptual objective speech quality metric to evaluate noise suppressors,''
  in \emph{Proc. IEEE ICASSP}, 2022, pp. 886--890.

\bibitem{POLQA-Beerends2013}
J.~G. Beerends \emph{et~al.}, ``Perceptual objective listening quality
  assessment ({POLQA}), the third generation {ITU-T} standard for end-to-end
  speech quality measurement part {I}--—temporal alignment,'' \emph{Journal
  of The Audio Engineering Society}, vol.~61, no.~6, pp. 366--384, 2013.

\bibitem{PESQ-Rix2001}
A.~W. Rix \emph{et~al.}, ``Perceptual evaluation of speech quality ({PESQ})---a
  new method for speech quality assessment of telephone networks and codecs,''
  in \emph{Proc. IEEE ICASSP}, vol.~2, 2001, pp. 749--752.

\bibitem{ESTOI-Jensen2016}
J.~Jensen and C.~H. Taal, ``An algorithm for predicting the intelligibility of
  speech masked by modulated noise maskers,'' \emph{IEEE/ACM Trans. ASLP.},
  vol.~24, no.~11, pp. 2009--2022, 2016.

\bibitem{SDR-Vincent2006}
E.~Vincent, R.~Gribonval, and C.~F{\'e}votte, ``Performance measurement in
  blind audio source separation,'' \emph{IEEE Trans. ASLP.}, vol.~14, no.~4,
  pp. 1462--1469, 2006.

\bibitem{MCD-Kubichek1993}
R.~Kubichek, ``Mel-cepstral distance measure for objective speech quality
  assessment,'' in \emph{Proceedings of IEEE Pacific Rim Conference on
  Communications Computers and Signal Processing}, 1993, pp. 125--128.

\bibitem{LSD-Gray1976}
A.~Gray and J.~Markel, ``Distance measures for speech processing,'' \emph{IEEE
  Transactions on Acoustics, Speech, and Signal Processing}, vol.~24, no.~5,
  pp. 380--391, 1976.

\bibitem{NISQA-Mittag2021}
G.~Mittag \emph{et~al.}, ``{NISQA}: A deep {CNN}-self-attention model for
  multidimensional speech quality prediction with crowdsourced datasets,'' in
  \emph{Proc. Interspeech}, 2021, pp. 2127--2131.

\bibitem{Evaluation-Pirklbauer2023}
J.~Pirklbauer \emph{et~al.}, ``Evaluation metrics for generative speech
  enhancement methods: Issues and perspectives,'' in \emph{Speech
  Communication; 15th ITG Conference}, 2023, pp. 265--269.

\bibitem{SpeechBERTScore-Saeki2024}
T.~Saeki \emph{et~al.}, ``{SpeechBERTScore}: Reference-aware automatic
  evaluation of speech generation leveraging {NLP} evaluation metrics,''
  \emph{arXiv preprint arXiv:2401.16812}, 2024.

\bibitem{Pushing-Jung2022}
J.-w. Jung \emph{et~al.}, ``Pushing the limits of raw waveform speaker
  recognition,'' in \emph{Proc. Interspeech}, 2022, pp. 2228--2232.

\bibitem{OWSMv3.1-Peng2024}
Y.~Peng \emph{et~al.}, ``{OWSM} v3. 1: Better and faster open {Whisper}-style
  speech models based on {E}-{B}ranchformer,'' \emph{arXiv preprint
  arXiv:2401.16658}, 2024.

\bibitem{Speech-Cohen2001}
I.~Cohen and B.~Berdugo, ``Speech enhancement for non-stationary noise
  environments,'' \emph{Signal Processing}, vol.~81, no.~11, pp. 2403--2418,
  2001.

\bibitem{Towards-Lu2022}
Y.-J. Lu \emph{et~al.}, ``Towards low-distortion multi-channel speech
  enhancement: The {ESPnet}-{SE} submission to the {L3DAS22} challenge,'' in
  \emph{Proc. IEEE ICASSP}, 2022, pp. 9201--9205.

\bibitem{ESPnet_SE-Li2021}
C.~Li \emph{et~al.}, ``{ESPnet-SE}: End-to-end speech enhancement and
  separation toolkit designed for {ASR} integration,'' in \emph{Proc. IEEE
  SLT}, 2021, pp. 785--792.

\end{thebibliography}

\begin{table*}
\caption{Breakdown results of different speech enhancement models on the non-blind test set of the URGENT challenge. \textbf{SF}: sampling frequency. \textbf{SNR}: signal-to-noise ratio. \textbf{RIR}: whether or not to apply the room impulse response. \textbf{Distortion}: type of additional distortions in the sample. Metrics with $\uparrow$ indicate the higher the better, while those with $\downarrow$ indicate the lower the better.}
\label{tab:breakdown}
\resizebox{0.52\textwidth}{!}{%
\begin{tabular}{l|ll}
\toprule
\multirow{2}{*}{\textbf{Models}} & \multicolumn{2}{c}{\emph{non-intrusive SE metrics}} \\
 & \textbf{DNSMOS OVRL $\uparrow$} & \textbf{NISQA MOS $\uparrow$} \\
\midrule
\rowcolor[HTML]{EEEEEE} 
Noisy input & 1.64 & 1.76\\
\textbf{SF}: 16 kHz / 22.05 kHz / 24 kHz / 32 kHz / 48 kHz & 1.83 / 1.72 / 1.58 / 1.61 / 1.26 & 1.62 / 1.72 / 1.84 / 1.74 / 1.77 \\
\textbf{SNR}: 0 dB / 5 dB / 10 dB / 15 dB & 1.34 / 1.63 / 1.86 / 2.04 & 1.40 / 1.79 / 2.01 / 2.19 \\
\textbf{RIR}: without / with & 1.84 / 1.43 & 2.01 / 1.50 \\
\textbf{Distortion}: none / bandwidth\_limitation / clipping & 1.69 / 1.69 / 1.53 & 2.03 / 1.83 / 1.41 \\
 &  & \\
\rowcolor[HTML]{F0F0F0} 
Conv-TasNet & 2.31 & 2.71 \\
\textbf{SF}: 16 kHz / 22.05 kHz / 24 kHz / 32 kHz / 48 kHz & 2.59 / 2.44 / 2.27 / 2.34 / 1.50 & 2.79 / 2.69 / 2.65 / 2.59 / 2.90 \\
\textbf{SNR}: 0 dB / 5 dB / 10 dB / 15 dB & 2.21 / 2.36 / 2.36 / 2.40 & 2.54 / 2.77 / 2.84 / 2.86 \\
\textbf{RIR}: without / with & 2.74 / 1.87 & 3.33 / 2.08 \\
\textbf{Distortion}: none / bandwidth\_limitation / clipping & 2.31 / 2.31 / 2.30 & 2.80 / 2.79 / 2.55 \\
 &  & \\
\rowcolor[HTML]{F0F0F0} 
BSRNN & 2.41 & 3.05 \\
\textbf{SF}: 16 kHz / 22.05 kHz / 24 kHz / 32 kHz / 48 kHz & 2.71 / 2.55 / 2.36 / 2.54 / 1.51 & 3.06 / 3.07 / 3.00 / 2.97 / 3.23 \\
\textbf{SNR}: 0 dB / 5 dB / 10 dB / 15 dB & 2.36 / 2.47 / 2.43 / 2.43 & 2.86 / 3.17 / 3.18 / 3.16 \\
\textbf{RIR}: without / with & 2.85 / 1.97 & 3.86 / 2.22 \\
\textbf{Distortion}: none / bandwidth\_limitation / clipping & 2.42 / 2.40 / 2.41 & 3.03 / 3.00 / 3.10 \\
 &  & \\
\rowcolor[HTML]{F0F0F0} 
TF-GridNet & 2.43 & 3.06 \\
\textbf{SF}: 16 kHz / 22.05 kHz / 24 kHz / 32 kHz / 48 kHz & 2.67 / 2.58 / 2.38 / 2.60 / 1.58 & 3.06 / 3.07 / 2.99 / 3.10 / 3.38 \\
\textbf{SNR}: 0 dB / 5 dB / 10 dB / 15 dB & 2.38 / 2.47 / 2.44 / 2.46 & 2.98 / 3.15 / 3.11 / 3.08 \\
\textbf{RIR}: without / with & 2.89 / 1.95 & 3.92 / 2.19 \\
\textbf{Distortion}: none / bandwidth\_limitation / clipping & 2.42 / 2.42 / 2.43 & 3.14 / 2.83 / 3.21 \\
\bottomrule
\end{tabular}
}%

\vspace{5pt}

\resizebox{\textwidth}{!}{%
\begin{tabular}{l|lllll}
\toprule
\multirow{2}{*}{\textbf{Models}} & \multicolumn{5}{c}{\emph{intrusive SE metrics}} \\
 & \textbf{PESQ $\uparrow$} & \textbf{ESTOI (×100) $\uparrow$} & \textbf{SDR (dB) $\uparrow$} & \textbf{MCD $\downarrow$} & \textbf{LSD $\downarrow$} \\
\midrule
\rowcolor[HTML]{EEEEEE} 
Noisy input & 1.63 & 70.40 & 6.11 & 6.76 & 3.99 \\
\textbf{SF}: 16 kHz / 22.05 kHz / 24 kHz / 32 kHz / 48 kHz & 1.59 / 1.65 / 1.64 / 1.70 / 1.60 & 70.39 / 71.05 / 71.50 / 68.07 / 64.52 & 5.58 / 6.08 / 6.50 / 5.59 / 5.75 & 8.01 / 6.58 / 6.22 / 7.09 / 6.66 & 4.94 / 4.09 / 3.53 / 4.19 / 3.66 \\
\textbf{SNR}: 0 dB / 5 dB / 10 dB / 15 dB & 1.24 / 1.56 / 1.89 / 2.24 & 55.38 / 73.14 / 81.79 / 86.98 & -0.05 / 6.52 / 10.64 / 13.76 & 9.32 / 7.49 / 6.31 / 5.34 & 4.28 / 4.01 / 3.81 / 3.56 \\
\textbf{RIR}: without / with & 1.47 / 1.79 & 69.65 / 71.17 & 4.27 / 7.97 & 9.97 / 5.14 & 4.57 / 3.40 \\
\textbf{Distortion}: none / bandwidth\_limitation / clipping & 1.78 / 1.72 / 1.40 & 73.48 / 72.34 / 65.37 & 8.11 / 7.51 / 2.68 & 6.92 / 8.12 / 7.68 & 2.77 / 5.76 / 3.45 \\
 &  &  &  &  & \\
\rowcolor[HTML]{F0F0F0} 
Conv-TasNet & 2.42 & 79.91 & 14.42 & 3.23 & 2.73 \\
\textbf{SF}: 16 kHz / 22.05 kHz / 24 kHz / 32 kHz / 48 kHz & 2.30 / 2.37 / 2.50 / 2.30 / 2.45 & 79.50 / 79.81 / 81.71 / 75.63 / 74.30 & 14.17 / 14.41 / 14.48 / 13.78 / 15.04 & 3.34 / 2.99 / 3.29 / 3.37 / 3.17 & 3.26 / 2.86 / 2.35 / 3.43 / 2.62 \\
\textbf{SNR}: 0 dB / 5 dB / 10 dB / 15 dB & 1.91 / 2.45 / 2.78 / 3.06 & 69.29 / 82.77 / 87.74 / 90.90 & 10.49 / 14.90 / 17.24 / 19.14 & 3.61 / 3.19 / 2.97 / 2.77 & 2.87 / 2.69 / 2.62 / 2.57 \\
\textbf{RIR}: without / with & 2.47 / 2.36 & 84.51 / 75.25 & 15.72 / 13.09 & 3.93 / 2.52 & 2.80 / 2.65 \\
\textbf{Distortion}: none / bandwidth\_limitation / clipping & 2.49 / 2.37 / 2.39 & 81.08 / 79.62 / 79.05 & 15.87 / 14.44 / 12.93 & 2.88 / 3.67 / 3.15 & 2.45 / 3.11 / 2.61 \\
 &  &  &  &  & \\
\rowcolor[HTML]{F0F0F0} 
BSRNN & 2.66 & 83.29 & 14.89 & 2.75 & 2.66 \\
\textbf{SF}: 16 kHz / 22.05 kHz / 24 kHz / 32 kHz / 48 kHz & 2.57 / 2.61 / 2.75 / 2.52 / 2.62 & 83.14 / 83.18 / 85.03 / 79.64 / 77.03 & 14.79 / 14.91 / 14.88 / 14.65 / 15.24 & 2.69 / 2.57 / 2.89 / 2.89 / 2.60 & 3.22 / 2.90 / 2.30 / 3.09 / 2.28 \\
\textbf{SNR}: 0 dB / 5 dB / 10 dB / 15 dB & 2.16 / 2.72 / 3.02 / 3.24 & 74.66 / 85.75 / 89.66 / 92.09 & 11.39 / 15.31 / 17.50 / 19.00 & 3.00 / 2.70 / 2.56 / 2.49 & 2.79 / 2.64 / 2.58 / 2.47 \\
\textbf{RIR}: without / with & 2.72 / 2.60 & 86.96 / 79.57 & 16.04 / 13.71 & 3.40 / 2.09 & 2.79 / 2.52 \\
\textbf{Distortion}: none / bandwidth\_limitation / clipping & 2.79 / 2.61 / 2.57 & 84.93 / 83.93 / 81.02 & 17.11 / 15.36 / 12.19 & 2.33 / 3.11 / 2.81 & 2.17 / 3.34 / 2.45 \\
 &  &  &  &  & \\
\rowcolor[HTML]{F0F0F0} 
TF-GridNet & 2.76 & 84.05 & 15.42 & 2.70 & 2.39 \\
\textbf{SF}: 16 kHz / 22.05 kHz / 24 kHz / 32 kHz / 48 kHz & 2.65 / 2.72 / 2.86 / 2.61 / 2.72 & 83.93 / 84.13 / 85.71 / 80.12 / 77.73 & 15.39 / 15.53 / 15.31 / 15.06 / 15.95 & 2.58 / 2.45 / 2.83 / 2.77 / 2.92 & 2.85 / 2.54 / 2.07 / 2.82 / 2.33 \\
\textbf{SNR}: 0 dB / 5 dB / 10 dB / 15 dB & 2.25 / 2.83 / 3.15 / 3.36 & 75.51 / 86.54 / 90.36 / 92.64 & 11.79 / 15.87 / 18.09 / 19.73 & 3.00 / 2.63 / 2.48 / 2.38 & 2.55 / 2.36 / 2.28 / 2.21 \\
\textbf{RIR}: without / with & 2.76 / 2.77 & 87.21 / 80.83 & 16.38 / 14.44 & 3.42 / 1.97 & 2.52 / 2.27 \\
\textbf{Distortion}: none / bandwidth\_limitation / clipping & 2.89 / 2.70 / 2.70 & 85.72 / 84.62 / 81.79 & 17.80 / 15.77 / 12.68 & 2.28 / 3.03 / 2.79 & 2.05 / 2.80 / 2.33 \\
\bottomrule
\end{tabular}
}%

\vspace{5pt}

\resizebox{0.915\textwidth}{!}{%
\begin{tabular}{l|ll|ll}
\toprule
\multirow{2}{*}{\textbf{Models}} & \multicolumn{2}{c|}{\emph{Downstream task independent metrics}} & \multicolumn{2}{c}{\textit{Downstream-task-dependent}} \\
 & \textbf{SpeechBERTScore $\uparrow$} & \textbf{Phoneme similarity $\uparrow$} & \textbf{Speaker similarity $\uparrow$} & \textbf{WAcc (\%) $\uparrow$} \\
\midrule
\rowcolor[HTML]{EEEEEE} 
Noisy input & 0.87 & 0.68 & 0.72 & 82.18 \\
\textbf{SF}: 16 kHz / 22.05 kHz / 24 kHz / 32 kHz / 48 kHz & 0.85 / 0.87 / 0.88 / 0.85 / 0.75 / 0.85 & 0.63 / 0.65 / 0.73 / 0.64 / 0.38 / 0.64 & 0.65 / 0.70 / 0.78 / 0.72 / 0.57 / 0.67 & 76.90 / 76.97 / 86.35 / 75.52 / 78.12 \\
\textbf{SNR}: 0 dB / 5 dB / 10 dB / 15 dB & 0.81 / 0.88 / 0.91 / 0.93 & 0.51 / 0.73 / 0.81 / 0.85 & 0.64 / 0.74 / 0.78 / 0.80 & 73.46 / 86.72 / 88.57 / 89.64 \\
\textbf{RIR}: without / with & 0.86 / 0.87 & 0.80 / 0.56 & 0.75 / 0.69 & 89.70 / 74.59 \\
\textbf{Distortion}: none / bandwidth\_limitation / clipping & 0.89 / 0.87 / 0.84 & 0.74 / 0.67 / 0.63 & 0.82 / 0.68 / 0.66 & 85.05 / 80.75 / 80.69 \\
 &  &  &  & \\
\rowcolor[HTML]{F0F0F0} 
Conv-TasNet & 0.85 & 0.73 & 0.70 & 76.82 \\
\textbf{SF}: 16 kHz / 22.05 kHz / 24 kHz / 32 kHz / 48 kHz & 0.82 / 0.82 / 0.87 / 0.84 / 0.85 & 0.67 / 0.70 / 0.79 / 0.65 / 0.70 & 0.58 / 0.69 / 0.76 / 0.70 / 0.70 & 69.99 / 71.37 / 81.51 / 68.39 / 74.53 \\
\textbf{SNR}: 0 dB / 5 dB / 10 dB / 15 dB & 0.78 / 0.86 / 0.89 / 0.92 & 0.61 / 0.78 / 0.82 / 0.85 & 0.59 / 0.73 / 0.78 / 0.81 & 64.80 / 82.51 / 85.77 / 87.59 \\
\textbf{RIR}: without / with & 0.89 / 0.80 & 0.85 / 0.61 & 0.73 / 0.66 & 86.04 / 67.51 \\
\textbf{Distortion}: none / bandwidth\_limitation / clipping & 0.86 / 0.84 / 0.84 & 0.77 / 0.70 / 0.73 & 0.76 / 0.63 / 0.70 & 79.35 / 73.52 / 77.56 \\
 &  &  &  & \\
\rowcolor[HTML]{F0F0F0} 
BSRNN & 0.87 & 0.80 & 0.77 & 82.53 \\
\textbf{SF}: 16 kHz / 22.05 kHz / 24 kHz / 32 kHz / 48 kHz & 0.85 / 0.85 / 0.89 / 0.87 / 0.87 & 0.76 / 0.77 / 0.85 / 0.76 / 0.76 & 0.71 / 0.76 / 0.82 / 0.76 / 0.77 & 76.27 / 77.84 / 86.64 / 76.70 / 80.30 \\
\textbf{SNR}: 0 dB / 5 dB / 10 dB / 15 dB & 0.82 / 0.88 / 0.91 / 0.93 & 0.72 / 0.83 / 0.86 / 0.88 & 0.70 / 0.80 / 0.83 / 0.85 & 74.29 / 86.61 / 88.75 / 89.63 \\
\textbf{RIR}: without / with & 0.92 / 0.82 & 0.90 / 0.70 & 0.81 / 0.74 & 89.98 / 75.01 \\
\textbf{Distortion}: none / bandwidth\_limitation / clipping & 0.88 / 0.86 / 0.87 & 0.84 / 0.78 / 0.79 & 0.84 / 0.71 / 0.77 & 84.98 / 80.34 / 82.23 \\
 &  &  &  & \\
\rowcolor[HTML]{F0F0F0} 
TF-GridNet & 0.87 & 0.81 & 0.78 & 82.87 \\
\textbf{SF}: 16 kHz / 22.05 kHz / 24 kHz / 32 kHz / 48 kHz & 0.85 / 0.85 / 0.89 / 0.88 / 0.88 & 0.76 / 0.78 / 0.85 / 0.77 / 0.77 & 0.70 / 0.76 / 0.82 / 0.77 / 0.78 & 77.30 / 78.15 / 86.78 / 76.16 / 81.13 \\
\textbf{SNR}: 0 dB / 5 dB / 10 dB / 15 dB & 0.82 / 0.89 / 0.91 / 0.93 & 0.72 / 0.84 / 0.86 / 0.89 & 0.70 / 0.80 / 0.83 / 0.85 & 74.70 / 87.02 / 88.95 / 89.90 \\
\textbf{RIR}: without / with & 0.92 / 0.83 & 0.90 / 0.71 & 0.81 / 0.74 & 89.54 / 76.14 \\
\textbf{Distortion}: none / bandwidth\_limitation / clipping & 0.89 / 0.87 / 0.87 & 0.85 / 0.78 / 0.79 & 0.85 / 0.70 / 0.78 & 85.74 / 80.82 / 82.02 \\
\bottomrule
\end{tabular}
}%
\end{table*} 
\end{document}